\documentclass[
reprint,
superscriptaddress,
showpacs,
amsmath,amssymb,
aps,
prb,
floatfix,
]{revtex4-1}

\usepackage{graphicx}
\usepackage{amsmath}
\usepackage{units}
\usepackage{amssymb}
\usepackage{color}
\usepackage{url}
\usepackage{bm}

\begin{document}

\title{Anomalous thermal decoherence in a quantum magnet measured \\
with neutron spin-echo spectroscopy}

\author{F. Groitl}
\altaffiliation[present address: ]{\'Ecole Polytechnique F\'ed\'erale de Lausanne, 1015 Lausanne, Switzerland and Paul Scherrer Institute, 5232 Villigen PSI, Switzerland}
\email{felix.groitl@psi.ch}
\affiliation{Helmholtz-Zentrum Berlin f\"ur Materialien und Energie GmbH, 14109 Berlin, Germany} 

\author{T. Keller}
\affiliation{Max Planck Institute For Solid State Research, 70569 Stuttgart, Germany}
\affiliation{Max Planck Society Outstation at the FRM II, 85748 Garching, Germany}

\author{K. Rolfs}%
\altaffiliation[present address: ]{Paul Scherrer Institute, 5232 Villigen PSI, Switzerland}
\affiliation{Helmholtz-Zentrum Berlin f\"ur Materialien und Energie GmbH, 14109 Berlin, Germany}

\author{D. A. Tennant}
\altaffiliation[present address: ]{Oak Ridge National Laboratory, TN 37831 Oak Ridge, USA}
\affiliation{Helmholtz-Zentrum Berlin f\"ur Materialien und Energie GmbH, 14109 Berlin, Germany}
\affiliation{Technische Universit\"at Berlin, Institut f\"ur Festk\"orperphysik, 10623 Berlin, Germany}

\author{K. Habicht}
\affiliation{Helmholtz-Zentrum Berlin f\"ur Materialien und Energie GmbH, 14109 Berlin, Germany}

\date{\today}

\begin{abstract}
The effect of temperature dependent asymmetric line broadening is investigated in Cu(NO$_3$)$_2\cdot$2.5D$_2$O, a model material for a 1-D bond alternating Heisenberg chain, using the high resolution neutron-resonance spin-echo (NRSE) technique. Inelastic neutron scattering experiments on dispersive excitations including phase sensitive measurements demonstrate the potential of NRSE to resolve line shapes, which are non-Lorentzian, opening up a new and hitherto unexplored class of experiments for the NRSE method beyond standard line width measurements. The particular advantage of NRSE is its direct access to the correlations in the time domain without convolution with the resolution function of the background spectrometer. This novel application of NRSE is very promising and establishes a basis for further experiments on different systems, since the results for Cu(NO$_3$)$_2\cdot$2.5D$_2$O are applicable to a broad range of quantum systems.
\end{abstract}


\maketitle

\section{introduction}

Thermal decoherence of quantum states is of wide-reaching importance for the 
application of quantum materials. The generic scenario encountered in condensed matter is due to quasiparticle interactions associated with exponential loss of coherence in the time domain and manifests in a symmetric Lorentzian-type line broadening in energy \cite{Bayrakci2006,Ronnow2001,Huberman2008,Xu2007}. 
According to the standard quasi-particle interaction theory the principal effect of the temperature is to increase this Lorentzian line width corresponding to shortening the lifetime by more frequent collisions. Such behavior is accessible in magnetic systems \cite{Forster1995,Marshall1971} where even the appearance of  universal line width behavior for chain systems has been proposed \cite{Sachdev2011,Zheludev2008,Kenzelmann2001}. However, in dimensionally constrained systems and those with hard core interactions it can be expected that strongly correlated effects should become evident and these will modify the decoherence in time away from an exponential form.

Experimental studies on Cu(NO$_3$)$_2\cdot$D$_2$O (copper nitrate), a model material for a 1-D bond alternating Heisenberg chain (AHC), have detected the development of non-Lorenzian line shapes using inelastic neutron scattering (INS) \cite{Tennant2012}. Non-perturbative calculations which take into account the cummulative effect of quantum interference effects in collisions have predicted such non-exponential decoherence in the time domain and their non-Lorentzian energy line shapes \cite{Essler2008,Essler2009}. Further, the direct application to the dimerized chain \cite{James2008,Goetze2010} provides approximate agreement with the temperature dependence observed experimentally. Subsequently non-Lorentzian-type line broadening has been observed in a 3-D dimerized magnet Sr$_3$Cr$_2$O$_8$, and it has been argued that these may be found in a broad range of quantum systems \cite{Quintero.PhysRevLett.109.127206,Tennant2012}.

\begin{figure}
\centering
\includegraphics[width=\columnwidth]{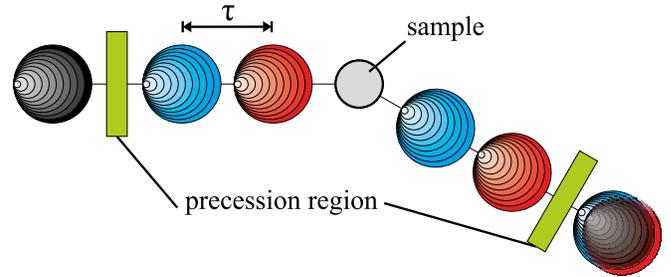}
\caption{Principle of spin echo measurements. In order to be sensitive to correlations in the sample the probing wavefunction itself must retain a well defined phase correlation. In standard neutron scattering instrumentation this correlation extends only over a single finite region in space (leftmost black area). With neutron spin echo the correlation volumes of the two spin states in the magnetic field are associated with different kinetic energies and therefore are spatially separated by the precession field before the sample (areas in red and blue mark correlation volumes of the up and down spin states respectively). They probe the same point in real space at different times. The spin echo time $\tau$ is the temporal separation of the correlation volumes probing the sample at times $t$ and $t+\tau$. The scattered wave functions preserve the time-separated correlation now including the imprint of sample-related time correlation. A second precession region reverses the separation. Finally, the expectation value of the neutron polarization $\left\langle\sigma_x\right\rangle$ is measured providing a sensitive probe of time-correlations in the sample \cite{HabichtNSEbook2003, Felber1998, Gaehler1998}.}
\label{fig:NRSE_sketch2}
\end{figure}

In the present work, we explore the neutron spin-echo triple-axis technique (NRSE-TAS) \cite{Keller2002,Habicht2004} as an alternative approach to the study of line shapes and widths of spin excitations. The particular advantage of the neutron spin-echo method is its capability to directly probe correlations in the time domain, the natural dimension for such processes. NRSE-TAS gives access to time scales which correspond to features in energy in the $\mu$eV-range and so is able to probe slow physical processes inaccessible to conventional INS. As a second advantage, background intensity has no influence on the line width or asymmetry due to the fact, that the background with its broad distribution in energy is depolarized and does not contribute to the measured NRSE signal. This is in contrast to conventional INS (ToF and TAS), where inaccuracies in background subtraction affects both the line width and the asymmetry. As a third advantage, the deconvolution of the data with the instrument resolution function necessary in conventional INS reduces to a simple normalization of the raw data in the case of spin-echo. 

The quantity measured by neutron spin-echo is the polarization of the neutron beam $P=\left\langle \sigma _{x}\right\rangle $, defined as the expectation value of the $x$-component of the neutron spin operator. An intuitive picture is given in terms of two correlation volumes associated with spin up and spin down spin states \cite{HabichtNSEbook2003, Felber1998} (Fig.\ref{fig:NRSE_sketch2}). The correlation volumes are finite spatial regions with a well-defined phase correlation of the neutron wavefunction. The longitudinal and transverse widths of the correlation volume are the inverse of the divergence and monochromaticity of the neutron beam, respectively. Inside the precession regions, the kinetic energies of the spin up and spin down states split\cite{Golub1993} and the two correlation volumes acquire a relative time delay $\tau$ (\emph{spin-echo time}). The spin states then scatter at the sample at times $t$ and $t+\tau$, where $\tau$ is identical to the van-Hove correlation time \cite{Gaehler1998}. After cancellation of the time-delay in a second inverted precession region, the scattered waves interfere at the detector such that the polarization $P$ is a direct measure of the time-dependence of the intermediate scattering function $I\left( Q,\tau \right) $: 
\begin{equation}\label{equ:polarizationandintermediate}
P=\left\langle \sigma_{x} \right \rangle \propto I\left( \bm{Q},\tau \right) e^{-i\omega_{0} \tau }+c.c. \\
\end{equation}
$ \hbar\omega _{0}$ is the mean energy of the excitation. The instrumental resolution is proportional to $\tau$, which is limited only by the homogeneity of the precession field and is independent of the size and shape of the correlation volumes. As theories are usually formulated in the momentum-energy space $(\bm{Q}, \omega)$, in practice eq.~\eqref{NRSEpolarization} is used to model spin-echo data: 
\begin{equation}\label{NRSEpolarization}
P = P_0(\tau) \int S(\bm{Q},\omega) T(\omega) \cos(\omega_0 \tau) d\omega,
\end{equation}
where $S(\bm{Q},\omega)$ is the scattering function. The factor $T(\omega)$ describes the transmission of the so-called background spectrometer, a TAS in the present case. Usually $T(\omega)$ is much broader than the narrow line widths studied with spin-echo and thus is set to unity. The resolution function $P_0(\tau)$ includes effects from the sample, such as mosaic spread and curvature of the dispersion sheet, and instrumental effects resulting, for example, from small imperfections of the precession regions \cite{Habicht2003}.

Classical NSE instruments based on DC fields generated by long solenoids were very successful during the past decades in the study of quasi-elastic scattering, i.e. with a mean energy transfer of zero. Famous examples of non-exponential relaxation with non-Lorentzian line shapes observed by NSE include relaxation of spin-glasses \cite{Mezei1980, Pickup2009}, the stretched exponential relaxation of glasses \cite{Mezei1987,Mezei21987}, deviations from the exponential relaxation in ferromagnets \cite{Mezei1986, Mezei1989} or the characteristic relaxation due to reptation in polymers \cite{Richter1990, Schleger1998}. For the study of excitations outside the quasi-elastic regime with finite energy, such as phonons and magnons, it proved necessary to combine the spin-echo and the triple axis (TAS) techniques and thus to select a small region in the $(\bm{Q}, \omega)$ space. A first version of the spin-echo TAS method using DC precession coils found indirect evidence for non-Lorentzian line shapes for a phonon mode in germanium\cite{Kulda2004}. In previous studies of spin excitations using NRSE-TAS, only Lorentzian line shapes were observed \cite{Bayrakci2006,Nafradi2011,Bayrakci2013}. 

\section{experiment}

Copper nitrate (Cu(NO$_3$)$_2\cdot$2.5D$_2$O) is a near ideal 1-D dimerized spin-$1/2$ antiferromagnet \cite{Garaj1968,Morosin1970,Xu2000,Tennant2003}. The alternating Heisenberg chains are formed of spin-1/2 moments on the Cu$^{2+}$ ions and symmetry equivalent chains lie along the $[1/2,1/2,1/2] $ and the $[1/2,-1/2,1/2] $ directions, which project onto the same direction on the (h 0 l)-plane. The dimerization gives rise to a singlet ground state and the elementary excitation is a triplet of spin-1 states \cite{Barnes1999} corresponding to excited dimer states that hop from site-to-site along the chain. For the dominant exchange couplings (interdimer $J=0.443\,$meV, intradimer $J^\prime=0.101\,$meV) the magnon bandwidth is small compared to the gap and due to the small alternation ratio $\alpha=J^\prime / J \approx 0.227$, there is a clear energy separation of about $0.5\,$meV between intra-band transitions, single magnon excitations, and two magnon continua even at high temperatures \cite{Tennant2012}.

\begin{figure} [hbt]
\centering
\includegraphics[width=\columnwidth]{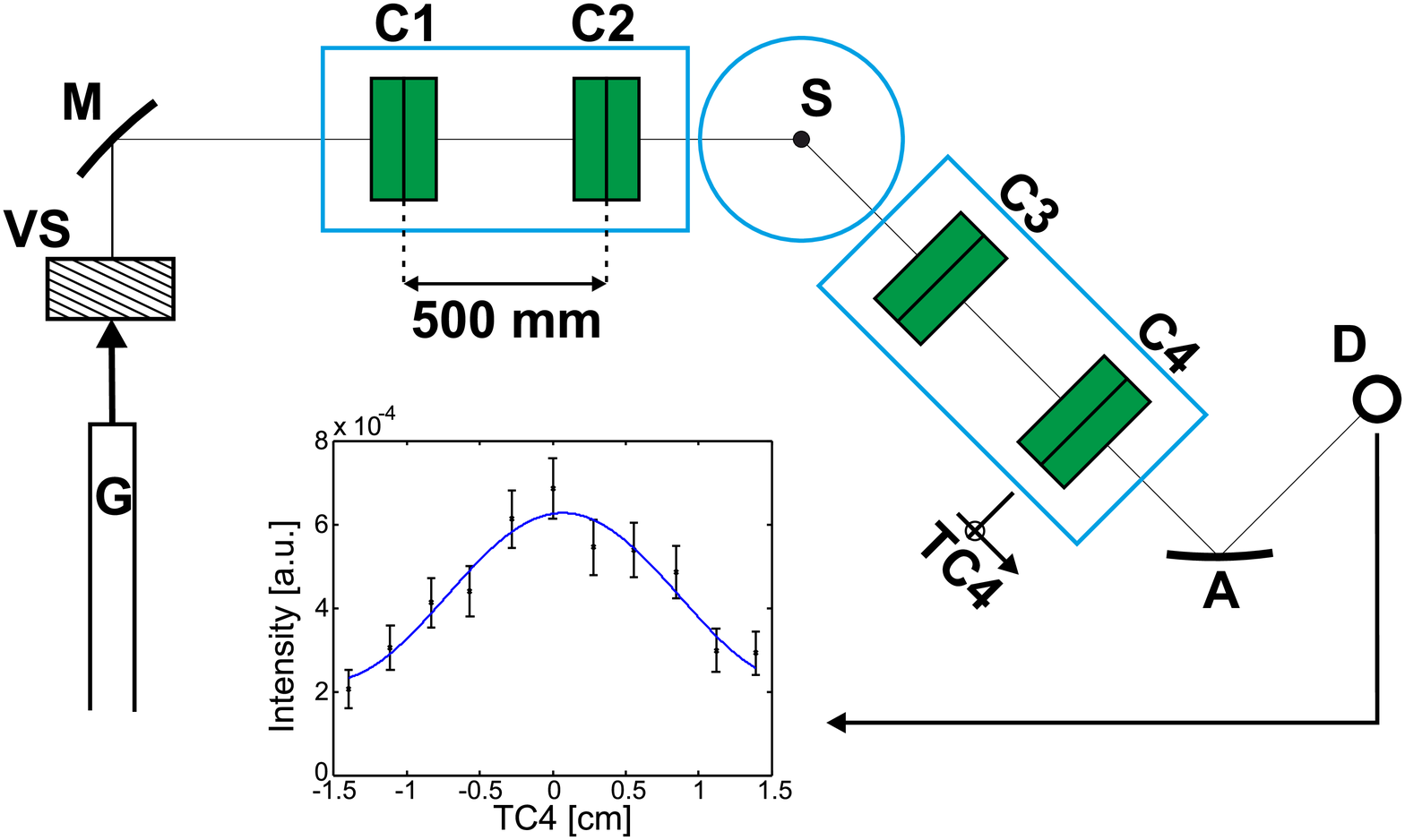}
\caption{TRISP spectrometer. \emph{M} and \emph{A} are the monochromator and analyzer,
as in TAS. \emph{S} is the sample and \emph{D} the detector; \emph{VS} indicates the
velocity selector. The resonance coil pairs (\emph{C1-C2} and \emph{C3-C4}) define the precession regions. The static fields $\bm{B_0}$ and the rf fields $\bm{B_1}$ are confined to the flipper coils. Mu-metal boxes and tubes (blue) enclosing the coils and the sample reduce external magnetic fields along the beam path to negligible values of $<5\,$mG to avoid spurious spin precession. Inset: Typical NRSE scan, detector counts \emph{vs.} position $TC4$ of the coil $C4$.}
\label{fig:CuN_Fig_TRISP_sketch}
\end{figure}

High-quality single crystals of Cu(NO$_3$)$_2\cdot$D$_2$O were grown at the HZB, Berlin, using the enrichment and solution growth method \cite{Tennant2012}. The single crystal used throughout the experiments had a mass of $4\,$g and a deuteration ratio of $>99.38\,$\%. 

The NRSE-TAS spectrometer TRISP \cite{Keller2002} (Fig. \ref{fig:CuN_Fig_TRISP_sketch}) at the FRM II was used to study the one-magnon mode at at the minimum of the dispersion corresponding to $\bm{Q}=(1\,0\,1)\,$r.l.u., $\hbar\omega_0=0.385\,$meV, where the intensity is maximized. The copper-nitrate crystal was aligned in the $(h\,0\,l)$ scattering plane in a closed cycle $^3$He cryostat. TRISP was operated with a graphite $(0\,0\,2)$ monochromator and a Heusler $(1\,1\,1)$ analyzer, with scattering sense $SM=-1$, $SS=-1$, $SA=1$ at the monochromator, sample, and analyzer, respectively ($-1$ is clockwise). With $k_i=1.7\,\text{\AA}^{-1}$, the TAS energy resolution in this configuration is $0.30\,\text{meV}$ (vanadium width, FWHM), which was in the present case sufficient to suppress the elastic background to about $2\,\%$ of the signal amplitude. The frequencies applied to the coils $C1-C4$ (see Fig. \ref{fig:CuN_Fig_TRISP_sketch}) were tuned according to the spin echo tuning conditions eqs.~(4,5) in ref.\cite{Habicht2004}. 

At TRISP, the polarization of the neutron beam (eq.~\eqref{NRSEpolarization}) is determined by scanning the coil \emph{C4} along the beam direction, such that the length of the second precession region differs by $TC4$ from the first one.  This leads to a sinusoidal variation of the count rate $I(\text{TC4})$, where one period $\Delta \text{TC4}=\hbar k_f/(m \nu_\text{eff})$ corresponds to a $2\pi$ rotation of the neutron spins. $\nu_\text{eff}$ is the effective neutron Larmor frequency\cite{Gaehler1988}, $m$ is the neutron mass. This scan is repeated for different values of $\tau$, with the present parameters $\tau[\text{ps}] = 0.145 \nu_\text{eff}\,[\text{kHz}]$. The polarization is the contrast of the modulated count rate 
\begin{equation}\label{eq:IvsTC4}
I(\tau, \text{TC4}) = I_0(1+P(\tau)\cos \left[\frac{2\pi (\text{TC4}-\text{TC4}_0(\tau))}{\Delta \text{TC4}(\tau)}\right]),
\end{equation}
where $I_0$ is the mean intensity corresponding to $P=0$, $\text{TC4}_0$ is a phase offset.

\begin{figure} [hbt]
\centering
\includegraphics[width=\columnwidth]{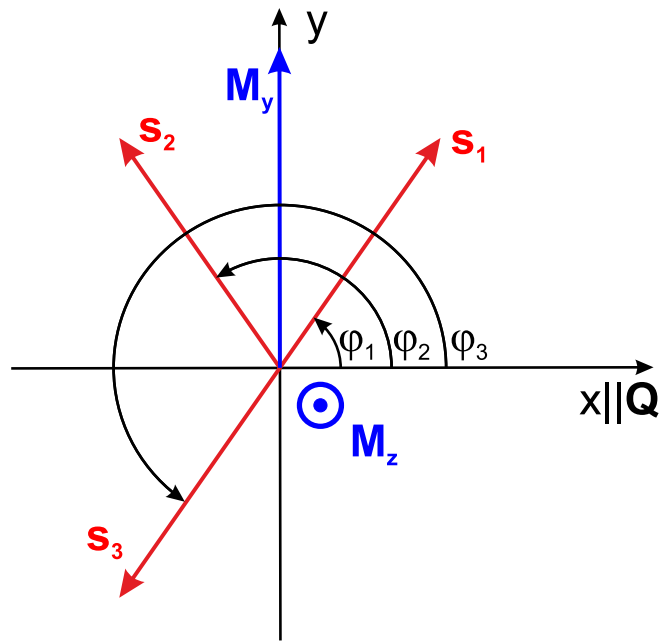}
\caption{Spin flip processes at the sample. The spins of the incident beam are spread within the horizontal xy-plane, where $x\parallel\bm{Q}$, $z$ is vertical. Only magnetic fluctuations $\bm{M}_y, \bm{M}_z \perp \bm{Q}$ contribute to the scattering cross section. The spin $\bm{s}_1$ with Larmor phase $\phi_1$ of the incident neutron is flipped to $\bm{s}_2$ or $\bm{s}_3$ by $\bm{M}_y$ or $\bm{M}_z$, respectively. The corresponding phases are $\phi_2 = \pi - \phi_1$ and $\phi_3 = \pi + \phi_1$.} 
\label{fig:FigSpinFlip}
\end{figure}

One complication met in spin-echo experiments on spin excitations arises from the $\pi$ spin flips of the neutron spins during the scattering process (Fig.~\ref{fig:FigSpinFlip}). In spin-echo, the neutron spins accumulate different Larmor phases in the first precession region due the spread in $k_i$, which is $\Delta k_i/k_i \simeq 0.015$ with $\bm{k_i}=1.7\,\text{\AA}$ in the present experiment. This spread of the phase is proportional to $\bm{B_0}$, and typically is much larger than $2\pi$ at the highest  fields. Thus at the sample, the neutron spin phases are spread within a plane perpendicular to $\bm{B_0}$, the \emph{precession plane}. At TRISP, $\bm{B}_0$ is vertical, and the precession plane is horizontal. This situation is different to 1-D polarization analysis\cite{Moon1969}, where at the sample all neutron spins are aligned in the same direction, parallel (or anti-parallel) to a guide field. If no spin flip occurs during scattering in the spin-echo spectrometer (nuclear scattering), the polarization is recovered in the second precession region with inverted field direction $-\bm{B}_0$ and forms the so called \emph{spin-echo} \cite{Mezei1980}. For the spin flip scattering in spin-echo experiments, the two cases of spin fluctuations $\bm{M}_y$ and $\bm{M}_z$ have to be distinguished (Fig \ref{fig:FigSpinFlip}): $\bm{M}_z$ adds a phase $\pi$ to the neutron spin phase, whereas $\bm{M}_y$ inverts the sign of the neutron spin phase, and thus effectively inverts the sign of the first precession field $\bm{B}_0$. Thus, to fulfill the echo condition for $\bm{M}_z$ ($\bm{M}_y$) fluctuations, the polarity of the fields must be anti-parallel (parallel)\cite{Tseng2015}. If both types of spin fluctuations contribute with equal amplitudes, it is better to choose the parallel field configuration, as in this case nuclear non-spinflip background is dephased and will not contribute to the polarization. In practice, the magnetic structure and the fluctuations are often not exactly known or obscured by the formation of domains. Thus at TRISP the suitable $\bm{B}_0$ configuration (parallel or anti-parallel) is experimentally determined. In the present case, both $\bm{M}_y$ and $\bm{M_z}$ were expected to contribute equally to the aforementioned $\pi$ flips, but the anti-parallel configuration showed better polarization close to the expected $P(\tau=0)=0.5$ and therefore was chosen for the subsequent measurements.

Spin-echo data were collected at 4 temperatures ($0.5\,\text{K}$, $2\,\text{K}$, $2.5\,\text{K}$, $3\,\text{K}$) below the characteristic activation temperature of the gap ($T=4.5\,$K), for $\tau$ in the range $14.5\,\text{ps}$ to $112.6\,\text{ps}$. Typical count rates were $10/\text{min}$ and $5.6/\text{min}$ at $0.5\,\text{K}$ and $3\,\text{K}$, respectively. A background rate of $1/\text{min}$ was subtracted from all scans.

\section{data analysis}

A phenomenological function describing an asymmetric modified Lorentzian lineshape has been used for the analysis of magnon line shapes measured by ToF-INS \cite{Tennant2012}:
\begin{equation}\label{equ:asym_model}
S\left(\omega\right)=\frac{1}{\pi}\frac{1}{1+\left(\omega/\Gamma - \alpha(\omega/\Gamma)^2+\gamma(\omega/\Gamma)^3\right)^2}
\end{equation}
Here, the argument of the usual Lorentzian is replaced by a polynomial, that includes two parameters to model the asymmetry, an asymmetry term $\alpha(\omega/\Gamma)^2$, and a damping term $\gamma(\omega/\Gamma)^3$. A symmetric Lorentzian is obtained for $\alpha, \gamma=0$. The function provides reasonable results for fitting the ToF data in the $\omega$ space, but tends to give unphysical solutions with several peaks at the zeros of the polynomial when applied to the spin echo data. To avoid these multiple peaks, we keep $\Gamma$ and $\alpha$ as independent variables, and calculate $\gamma$, such that the second derivative of $\partial^2S(\omega)/\partial\omega^2$ has only two zeros, defining the inflection points of the peak. Other representations of asymmetric Lorentzians\cite{Stancik2008} with only one asymmetry parameter gave no satisfactory description of the ToF data.

The usual way to model NRSE data consists in a 2-step fitting process. First, the magnitude of the polarization $P$ and the phase $\text{TC4}_0$ for one  $\tau$ value are determined by fitting eq.~(\ref{eq:IvsTC4}) to the raw data (Fig. \ref{fig:CuN_Fig_TRISP_sketch}, inset). Then the model $S(\omega)$ is fitted to $\left|P(\tau)\right|$ using eq.~\eqref{NRSEpolarization}. The phase $\text{TC4}_0$ in eq.~(\ref{eq:IvsTC4}) is usually neglected, as it contains only a trivial factor $\omega_0 \tau$. In the present case with asymmetric linewidths, $\text{TC4}_0$ also carries information about the asymmetry, and thus was included in the analysis. Besides  the structure factor $S(\omega)$, and the spectral width $\Delta k_i$ of the incident beam, also the aforementioned $\pi$ spin flips and the TAS energy resolution have to be modeled in the data analysis, but it proves difficult to implement all these factors in an analytical expression. In a first attempt we calculated the polarization $P(\tau, \text{TC4}, \omega_0, \Gamma, \alpha, \gamma, k_i, \Delta k_i)$ using a \emph{Monte Carlo (MC)} ray tracing simulation of the spectrometer, which tracks the spin phase of individual neutrons running through the precession regions and spectrometer. The TAS was simplified by defining a Gaussian spectrum for the incident neutrons ($\Delta k_i/k_i = 0.015$ (FWHM)), and a Gaussian probability distribution of energy transfers $\omega$ ($0.3\,\text{meV}$ FWHM). The fitting procedure based on this simulation took excessively long time to converge, as the minimization algorithm of the fitting function \cite{James1994} is disturbed by the statistical noise of the MC algorithm. Finally we simplified the model by using discrete equally spaced values instead of random numbers to select the $k_i$ and $\omega$.
\begin{figure} [hbt]
\centering
\includegraphics[width=\columnwidth]{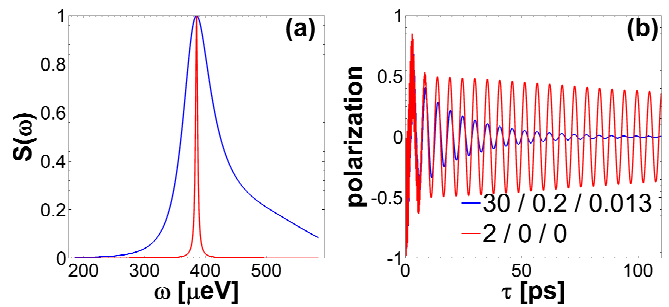}
\caption{(a) Asymmetric $S(\omega)$ according to eq.~(\ref{equ:asym_model}) with parameters $\Gamma [\mu eV] / \alpha / \gamma = 30/0.2/0.018$ (blue) and symmetric Lorentzian $2/0/0$ (red). (b) Calculated polarization $P(\tau, \text{TC4}=0)$ using the model described in the text. }
\label{fig:Fig_asym_linewidth}
\end{figure}

The model $P(\tau)$ is shown in Fig. \ref{fig:Fig_asym_linewidth} for a narrow symmetric Lorentzian and a broad asymmetric line. $P$ shows fast oscillations for $\tau<10\,ps$ arising from an interference of the spins flipped by fluctuations $\bm{M}_y$ and $\bm{M}_z$, which were assumed to contribute with equal weight. In the chosen anti-parallel $\bm{B}_0$ configuration the $\bm{M}_y$ component does not obey the echo condition and is rapidly damped. In the present experiment with $\tau_\text{min}=14.5\,ps$, these fast interference oscillations are not visible. The weight of the component $\bm{M}_y$ determines the constant factor $P_0$ in eq.~(\ref{eq:IvsTC4}), where for isotropic fluctuations we expect $P_0=0.5$, which is in good agreement with $P_0=0.47(3)$ obtained for the present data. The phase shift between the two curves in Fig.~\ref{fig:Fig_asym_linewidth}(b) is in the order of a few $10^\circ$ and is similar to the $3\,K$ data in Fig.~(\ref{fig:phaseshift_plot}).

\section{results and discussion}

\begin{figure}[hbt]
\centering
\includegraphics[width=\columnwidth]{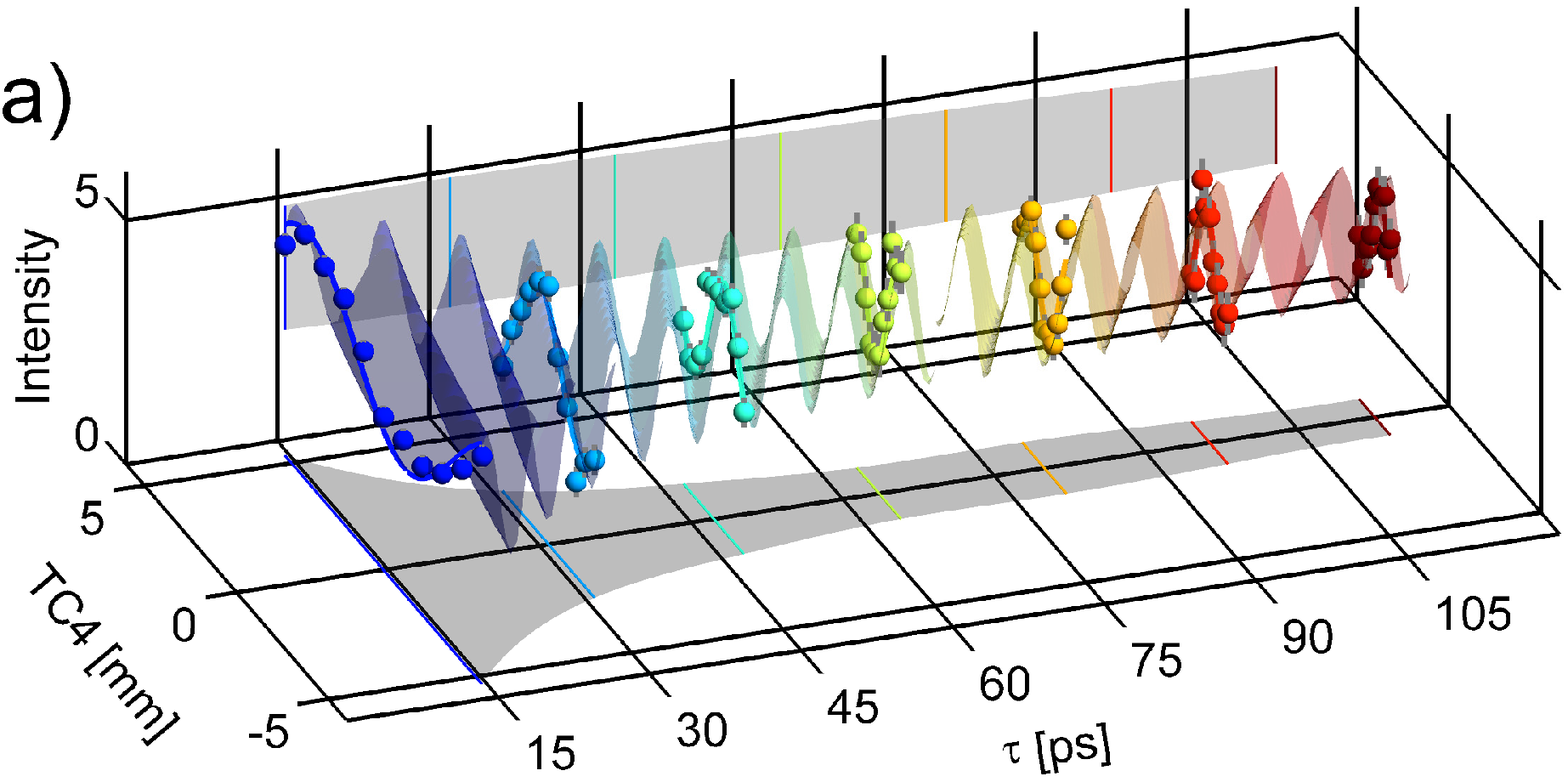}
\includegraphics[width=\columnwidth]{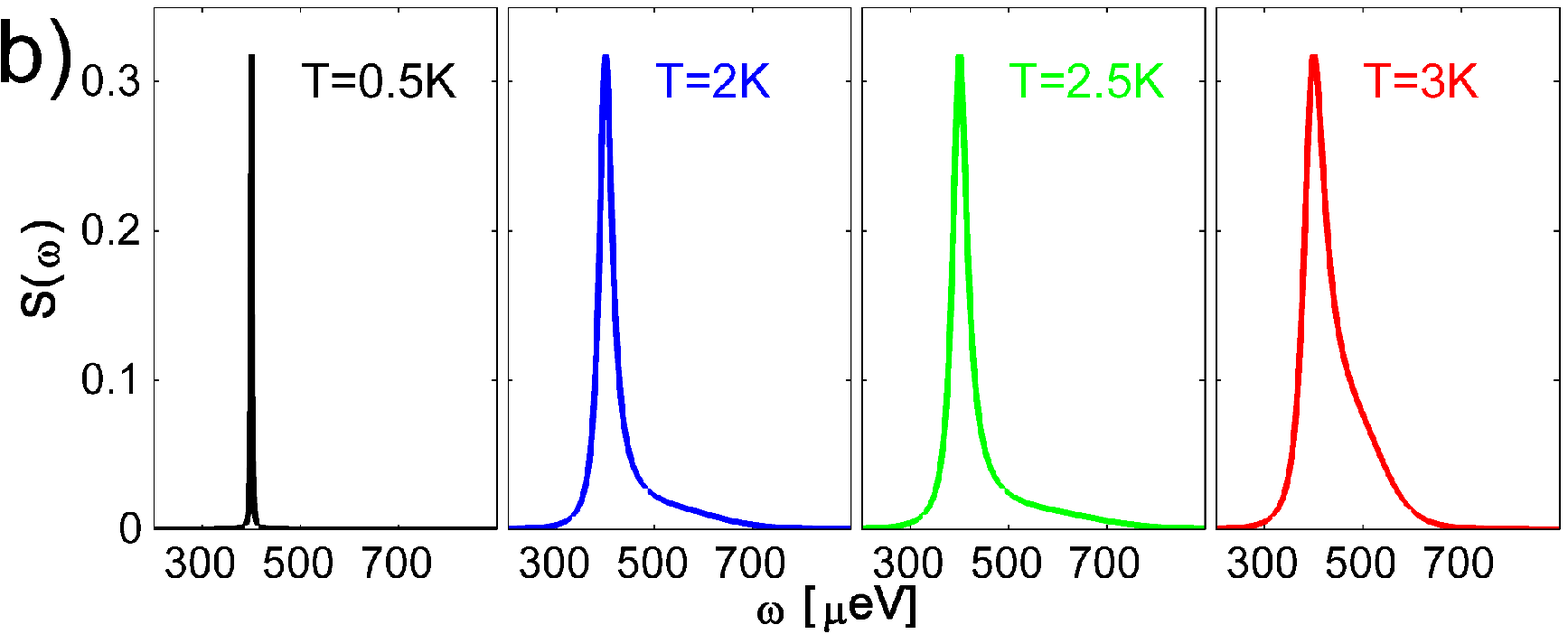}
\caption{Data (spheres) $I(\tau, \text{TC4}$) for $T=0.5\,K$. The ranges of the individual scans for different spin-echo times $\tau$ are projected to the bottom, while the intensity is projected to the side. The phenomenological model described in the text is fitted simultaneously to all scans of one temperature set. Hence, the phase, which carries additional information about the asymmetry, is considered. The fit of the model is shown as a semitransparent surface through all data points. \\
(b) The resulting fit parameters for the different temperature sets correspond to the plotted $S(\omega)$. With increasing temperature a clear asymmetry develops.} 
\label{fig:results3D}
\end{figure}

\begin{figure}[hbt]
\centering
\includegraphics[width=0.5\textwidth]{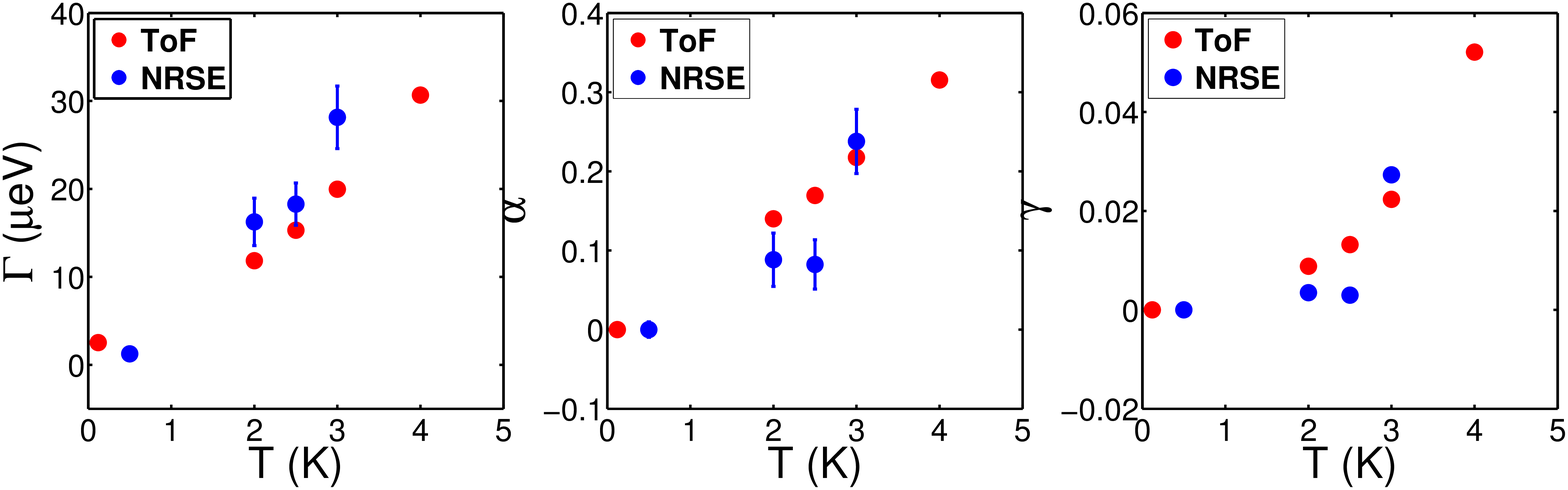}
\caption{Parameters resulting from a fit of the phenomenological model described in the text. This model was applied to the spin echo data (black) and the ToF data from ref. \cite{Tennant2012} (red).}
\label{fig:compare_asymparas}
\end{figure}

Fig.~\ref{fig:results3D} shows the data and the model $I(\tau, \text{TC4})$ for $T=0.5\,\text{K}$. The data where collected by scanning the length $TC_4$ of the second precession region for a set of spin-echo times $\tau$. The structure factors $S(\omega)$ resulting from the fit show a clear asymmetry increasing with $T$. The resulting parameters $\Gamma$, $\alpha$, and $\gamma$ (eq.~\eqref{equ:asym_model}) are plotted in Fig.~\ref{fig:compare_asymparas} and are compared to the parameters obtained for the ToF data fitted with the same model, where both for the spin-echo and the ToF data $\gamma$ is calculated from $\Gamma$ and $\omega$. The agreement between the two methods is surprisingly good, although the analysis of the ToF data included subtraction of a sloping background and deconvolution with a Gaussian resolution function ($\sigma = 0.017\,\text{meV}$). At first sight, the quality of the spin-echo data looks worse than ToF with larger errorbars and thus an increased scatter. On the other hand, the intrinsic width $\Gamma$ and the asymmetry $\alpha$ are obtained by spin-echo method without including assumptions about the background or the spectrometer resolution. The larger error bars on the spin-echo data are due to the low count rates, as in spin-echo the losses in the neutron polarizer and analyzer cost about 2/3 of the intensity, and additionally the polarization and thus the signal is reduced by a factor of 2 due to the $\pi$ flips of the neutron spins upon scattering. 

\begin{figure}[hbt]
\centering
\includegraphics[width=0.5\textwidth]{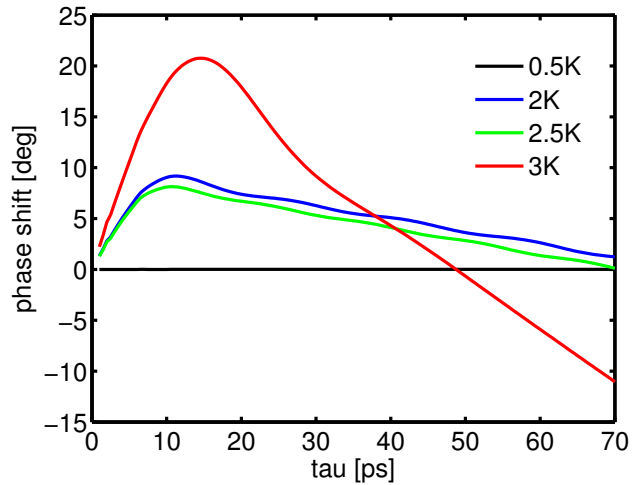}
\caption{Phase shift $\Delta\phi(\tau,T)$ of the polarization (Fig.~\ref{fig:Fig_asym_linewidth}) relative to $T=0.5K$.}
\label{fig:phaseshift_plot}
\end{figure}

\begin{figure}
\centering
\includegraphics[width=0.5\textwidth]{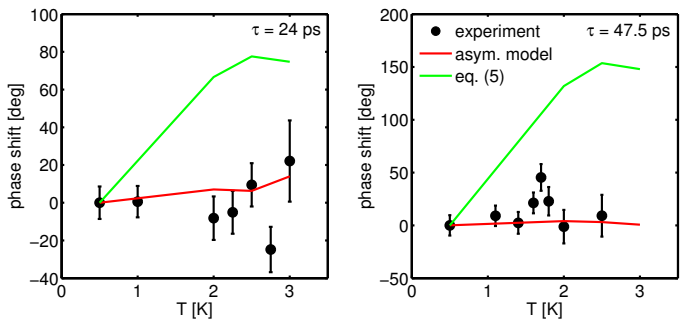}
\caption{Phase of the polarization \emph{vs.}~$T$ (black) from phase sensitive measurements at the dispersion minimum $Q=$(1.11 0 0.855) r.l.u., $E=0.385\,$meV for spin echo time $\tau=24.01\,$ps (left) and $47.57\,$ps (right). The data are compared to the phase shift calculated from the model discussed in the text (red) and from eq. \eqref{eq:phaseSymmetric} (green). }
\label{fig:phase_shift_sep12}
\end{figure}

In the following paragraphs the phase of the polarization resulting from our model (Fig.~\ref{fig:Fig_asym_linewidth}) is discussed. For a symmetric $S(\omega)$, a phase shift $\Delta\phi$ at a constant $\tau$ occurs only if the peak $\omega_0$ of the line shifts, but is otherwise independent of the width or shape of this line:
\begin{equation}
\label{eq:phaseSymmetric}
\Delta\phi_\text{sym} = \Delta\omega_0\times\tau
\end{equation}
Eq. (\ref{eq:phaseSymmetric}) is frequently used to determine the $T$-dependence of the excitation energy $\omega_0(T)$. With $\tau=100\,\text{ps}$ and a typical statistical error in the phase of a few degrees, energy shifts $\hbar\Delta\omega_0$ on the order of $1\mu\,\text{eV}$ can be detected without knowledge of the line shape and width \cite{Fak.2012}. \\
This powerful method for measuring the renormalization of $\omega_0$ fails in the present case with a $T$-dependent asymmetry of the line shape. The phase shift $\Delta\phi(\tau,T)=\phi(\tau,T)-\phi(\tau,T=0.5)$ between the polarization curves $P(\tau, \text{TC4}=0)$ (see Fig.~\ref{fig:Fig_asym_linewidth}) for the parameters obtained from the fits of the spin-echo data from Fig.~\ref{fig:results3D} are plotted in Fig.~\ref{fig:phaseshift_plot}. Here the phase of the lowest temperature $T=0.5\,K$ was taken as a reference. The clear non-linear evolution of the phase shift is a direct consequence of the non-linear lineshapes. This also means that applying eq.~\eqref{eq:phaseSymmetric} to non-symmetric lineshapes leads to meaningless results for the energy shift $\Delta\omega$. The relation between $\Delta\phi(\tau)$ and the asymmetry depends on the specific line shape and can in general only be calculated numerically from eq. \ref{NRSEpolarization}.\\
To further investigate the breakdown of the simple relation in eq.~\eqref{eq:phaseSymmetric}, we measured $\Delta\phi(T,\tau)$ for two values of $\tau$, $24\,\text{ps}$ and $47.5\,\text{ps}$, again such as all previous measurements at the minimum of the dispersion curve at $Q=(1.11\;0\;0.855)$. The data $\Delta\phi$ \emph{vs.}~$T$ are displayed as black dots in Fig.~\ref{fig:phase_shift_sep12}. The phase shift resulting from our model for the parameters of Fig.~\ref{fig:compare_asymparas} shows good agreement with the data within statistical accuracy. In contrast, the phase shift calculated from eq.~\eqref{eq:phaseSymmetric}, where $\Delta\omega$ was taken as the shift of the center of gravity of the asymmetric line with parameters Fig.~\ref{fig:compare_asymparas}, clearly disagrees with the experimental data.\\
The observation that the phase of the polarization obtained from asymmetric lineshapes with a T-dependent asymmetry is not proportional to $\Delta\omega$ is one of the main results of this paper. Thus in this case application of eq.~\eqref{eq:phaseSymmetric} to calculate $\Delta\omega$ will lead to wrong results. On the other hand the phase carries information about the line shape and thus should be included in the fit determining $S(\omega).$

\section{conclusions}

A new application of the NRSE method dedicated to line shape analysis on dispersive excitations has been presented opening up opportunities to look at strongly correlated quantum systems and extend beyond the more conventional systems investigated so far. This method was applied to the temperature dependent asymmetric line broadening present in the 1-D bond alternating Heisenberg chain material Cu(NO$_3$)$_2\cdot$D$_2$O. The results clearly demonstrate that the NRSE approach has the potential to detect anomalous effects due to strong correlations, which arise from the dimensional constraint of the system and hard core interactions of the excited states \cite{Tennant2012}. The particular advantage is the direct access to decoherence in the time domain and therefore the method complements the frequency measurements using conventional neutron spectroscopy. Further, the NRSE method does not depend on systematic errors arising from the convolution with the resolution function as in conventional spectroscopy. Its extremely high resolution also opens up the possibility to overcome the resolution limitations of energy domain measurements imposed by monochromating components. 

Furthermore, as the second important result it could be shown, that for a line shape differing from a Lorentzian the phase of the spin-echo signal becomes a non-linear function of the spin-echo time $\tau$. In this regard analysis beyond the conventional phase shift of a single spin echo time is necessary.

The results and applications are promising and establish a basis for further experiments on different quantum systems and high resolution measurements of edge singularities can now be made to test field theories and critical exponents in quantum critical systems.
\begin{acknowledgments}
The authors would like to thank Prof. Dr. Bella Lake (HZB) and Dr. Diana Lucia Quintero-Castro (HZB) for fruitful discussions and Kathrin Buchner (MPI) for technical support during the experiments. This work is based upon experiments performed at the TRISP instrument operated by MPG at the Forschungs-Neutronenquelle Heinz Maier-Leibnitz (FRM II), Garching, Germany. T.K. acknowledges financial support from the Deutsche Forschungsgemeinschaft through TRR80.
\end{acknowledgments}

\bibliography{CUN_06072015_V3}
\end{document}